\documentclass[11pt,a4paper]{article}
\usepackage[nohead, nomarginpar, margin=1in, foot=.25in]{geometry}

\usepackage{graphicx}

\usepackage{amsmath}

\usepackage[T1]{fontenc}
\usepackage[utf8]{inputenc}
\usepackage{authblk}

\usepackage[LGRgreek]{mathastext}

\usepackage{cite}
\date{}

\def\@listcomma@comma{\@ifnum{\@tempcnta>\tw@}{,}{}}

\begin{document}
\makeatletter
  \def\title@font{\Large\bfseries}
  \let\ltx@maketitle\@maketitle
  \def\@maketitle{\bgroup%
    \let\ltx@title\@title%
    \def\@title{\resizebox{\textwidth}{!}{%
      \mbox{\title@font\ltx@title}%
    }}%
    \ltx@maketitle%
  \egroup}
\makeatother

\title{Azimuthal correlations of D-mesons in $p$+$p$ and $p$+Pb collisions at LHC energies}

\author[1,\footnote{younus.presi@gmail.com}] {M.Younus}
\author[1,2] {S.K.Tripathy}
\author[1] {P.K.Sahu}
\author[2] {Z. Naik}

\affil[1] {Institute of Physics, Bhubaneswar-751005, India}
\affil[2] {Sambalpur University, Burla-768019, India}

\maketitle

\begin{abstract}
We study the correlations of D mesons produced in $p$+$p$ 
and $p$+Pb collisions. These are found to be sensitive to the effects of the cold nuclear medium and the 
transverse momentum ($p_T$) regions we are looking into. In order to put this on a quantitative footing, as a first step we analyse the azimuthal correlations of  D meson-charged hadron(Dh), and then predict the same for D meson -anti D meson ($D\overline{D}$) pairs in $p$+$p$ and $p$+Pb collisions with strong coupling at leading order $\cal{O}$($\alpha_{s}^{2}$) and next to leading order
 $\cal{O}$($\alpha_{s}^{3}$) 
which includes space-time evolution (in both systems), as well cold nuclear matter effects (in  $p$+Pb). 
This also sets the stage and baseline for the identification and
study of medium modification of azimuthal 
correlations in relativistic collision of heavy nuclei at the Large Hadron
Collider.
\end{abstract}

\section{Introduction}
\label{intro}
With the accumulation of data from Large Hadron Collider experiments, we 
have several published experimental results indicating formation of hot 
and dense nuclear matter commonly called quark gluon plasma(QGP)\cite{qgp1, qgp1_a, qgprev1,qgprev1_a}. Various 
signatures like jet quenching, elliptic flow etc. have broadly established the possible 
evidences for QGP formation in ultra-relativistic heavy ion collision. 
However the systematic analysis of data will continue for another couple of years to draw firm conclusion.
On the other hand, fixing the baseline with 
$p+p$ collisions for $Pb+Pb$ collisions is important. 
In ALICE-LHC,  $p+p$ collision experiments have become source of many new physics studies, even though QGP is not believed to be formed in such collisions.

And effects such as nuclear shadowing and initial cold nuclear matter scattering, which are generally overshadowed by hot and dense nuclear matter effects, are finding prominence in recent calculations on initial state nuclear effects and have been found to affect the final particle spectra. 

This has led to experiments such as $d$+Au 
and $p$+Au experiments at RHIC, BNL and $p$+Pb experiments at LHC,CERN experiments, whose up-to-date 
analysed data show considerable effect due to initial cold nuclear matter(CNM) effect on $p_T$ 
spectra on final hadrons~\cite{cnmexp1,cnmexp1_a}.

Alongside jets and photons\cite{suppress1, suppress1_a, suppress1_b, suppress2, suppress2_a, suppress2_b, exp1, exp1_a}, heavy quarks too are used to probe the QGP as they 
(only charm and bottom quarks are considered in general) offer several
unique advantages. The conservation of heavy flavour in strong interaction
 ensures that they are produced in pairs $(Q\overline{Q})$ only. 
The mass of heavy quark $(M_c = 1.5 \ GeV, M_b = 4.5 \ GeV)$ suggests that, large momentum transfer $(Q^2 >> \Lambda_{QCD}^{2})$ is necessary for their production. And thus one may use pQCD techniques for calculating heavy quark cross-section.
Heavy quarks' large masses also ensure that the heavy mesons would stand out among in-numerous pions. 
Their large mass also limits their production to the pre-equilibrium 
phase of heavy ion collisions, whereas production from other phases does not add much to their 
cross-sections~\cite{prod1,prod1_a,prod1_b}. 

Heavy quarks are buffeted by light quarks and gluons during their passage through the
quark gluon plasma, and even though they lose energy and momentum via drag and diffusion 
substantially~\cite{drag,drag_a,drag_b, drag_c,diffusion,diffusion_a,diffusion_b}, their direction of motion may not change considerably. 
This should make them a valuable probe for the properties of the initial nuclear effects, 
pre-equilibrium dynamics and QGP,  which also depend 
on the reaction plane. It is also not yet established that heavy
quarks will completely thermalize in the plasma formed at RHIC 
and LHC energies ~\cite{hvqtherm1,hvqtherm1_a}. Thus, the azimuthal correlation of heavy quarks integrated
over $p_T$ may be reasonably immune to the energy loss suffered by them and 
carry information on initial nuclear effects and geometry.

Here we discuss azimuthal correlations of heavy quarks ~\cite{corr1,corr1_a,corr2,corr2_a,corr3,corr3_a} in brief. 
Consider charm quarks (say) produced from the primary processes in $p+p$ collisions, 
$gg \rightarrow Q\overline{Q}$ at leading order and 
$gg \rightarrow gQ\overline{Q}$ at next-to-leading order.
In the absence of any intrinsic $k_T$ for partons, the quarks from the first process
will be produced back-to-back,
while those from the second process will not only have back-to-back and collinear correlation 
but also throughout $\Delta\phi$ range due to additional
accompanying recoiling gluons, which renders a small $k_T$ to heavy quark pair. 
And  there is a presence of further $k_T$ broadening 
due to multi-parton scattering, which may push the correlating pair more towards 
the collinearity~\cite{rsy1}. The recoiling gluons associated with a 
heavy quark will also form a correlation in azimuthal angle and are affected by the 
initial cold nuclear matter effects. 
We realise that in addition to this picture, the splitting $g \rightarrow Q\overline{Q}$, 
will produce collinear heavy quarks and almost back-facing associated gluon, 
while the process $gg \rightarrow Q\overline{Q}g$, where
a gluon is radiated by one of the heavy quarks at NLO will essentially give rise to a 
flat or broadened azimuthal correlation in both $D-\overline{D}$ and $D-h$.

At the final stage the charm quarks and associated gluons will fragment into 
$D$ mesons and charged hadrons ($\pi^{\pm},K^{\pm},p^{\pm}$) respectively and provide the experimentally 
observable $D-h$ and $D-\overline{D}$ correlations. Inclusion of 
$k_T$ broadening effect due to multi-parton scattering and nuclear shadowing on these correlations in 
the case of $p$+Pb collisions can be discerned using phenomenological 
models such as HIJING, AMPT and NLO-pQCD. Also a space-time evolution 
of charm quarks and its associated partons will also have effects on 
final observed $D-h$ and $D-\overline{D}$ azimuthal spectra, which will be discussed 
in the following sections. A comparison of the energy loss
suffered by the recoiling parton and the heavy-quarks in the $Pb+Pb$ scenario, will also allow
us to obtain flavour dependence of the energy loss. 
We also realise that $R_{AA}$ and $R_{pA}$ are not fully able to 
discriminate between different mechanisms of energy loss and initial cold nuclear matter 
effects  and the correlations of the leading heavy mesons and associated hadrons are slowly emerging as 
more discerning probes. As we know NLO-pQCD results are easily approximated
by multiplying the results for LO pQCD with a $K$ factor ~\cite{jamiln}.
Any initial nuclear effects and shadowing will have 
consequences on such correlation spectra. Thus one may study any deviations of $p+$Pb 
from $p+p$ collisions to obtain results
due to cold nuclear matter (CNM) modification and we will have a qualitative understanding of $p+p$ and $p+Pb$ collisions
before we can accurately decipher the $Pb$+$Pb$ collisions.

The present work aims to investigate azimuthal correlation, of heavy quark-anti quark pairs and 
heavy quark-associated parton via $D-\overline{D}$ and $D-h$ correlations, in $p+p$ and $p+$Pb collisions 
at $\sqrt{s}$=7 TeV and $\sqrt{s_{NN}}$=5.02 TeV respectively. 
This also sets the stage for the study of the deviations in these due to medium modifications
in heavy ion collisions at the corresponding energies.

The paper is organised as follows. In the next section, we discuss various
mechanisms to calculate azimuthal correlations for $p+p$ and $p+$Pb collisions using  HIJING, AMPT and NLO-pQCD.
Our results for $p+p$ and $p+$Pb collisions 
are discussed in Section \ref{result}, followed by conclusion in Section \ref{summary}.

\section{Models used}
\label{Models_used}

\subsection{The HIJING model}
\label{Hijing}
Heavy Ion Jet INteraction Generator (HIJING) \cite{hijing} is a two component (hard and soft) 
Monte Carlo program for hadron production in p+p, p+A and A+A collisions.
 Depending on transverse momentum exchange, the soft component is guided by Lund FRITIOF \cite{lund_fritiof,lund_fritiof_a} and
Dual Parton Model \cite{dpm,dpm_a,dpm_b} and modelled by formation of strings; whereas  PYTHIA \cite{pythia,pythia_a} 
deals with hard and semi-hard interactions which lead to the formation of energetic minijet partons.
 
Gluon productions from both the processes are included as kinks in the strings.  Excited strings are 
also assumed to have soft gluon radiation induced by the soft interactions. Finally, excited stings 
are fragmented into hadrons according to Lund fragmentation scheme \cite{lund_jet_model,lund_jet_model_a}.\\
This can be written as follows:
\begin{equation}
f(z) \approx z^{-1} (1-z)^a exp[-\frac{b(m^2+p_t^2)}{z}],
\end{equation}
where z is the light-cone momentum fraction, m and $p_{t}$ are mass and momentum of string respectively, a and b are Lund parameters.

Cross section for the hard process in the leading order (LO), 
can be written as follows:
\begin{equation}
\frac{d\sigma_{jet}}  {dp_{T}^{2} dy_1 dy_2} =   K \sum_{a,b} x_1 f_a(x_1,p_T^2)  x_2 f_b(x_2,p_T^2) 
 \times \frac{d\hat{\sigma}_{ab(\hat{s},\hat{t},\hat{u})}} {d\hat{t}}, 
\end{equation}
where $a,b$ are the parton species, $y_1, y_2$ are the rapidities of the scattered partons, $x_1, x_2$ are the fractions of momentum carried by the initial partons,
$\sigma_{ab}$ is parton-parton cross-section and s, t, u are standard Mandelstrem variables. 
A value of K = 2.0 has been used to account for next to leading order (NLO) corrections to cross section. The
parton structure function  $f_a(x, Q^2)$ is the Duke-Owens structure function (set-1) \cite{duke_own}.

To explain cold and hot nuclear effects, mass dependence of shadowing effect on parton structure 
function \cite{shadow_massdepend,shadow_massdepend_a} , EMC effect \cite{emc,emc_a} and 
effective energy loss of high $p_{T}$ jets \cite{energy_loss,energy_loss_a}(heavy ion scenario), are also used in HIJING.  

The parametric form of shadowing can be expressed as follows:
\begin{align}
R_{A}(x) \equiv \frac{f_{a/A}(x)}  {A f_{a/N}(x)}  = 1+1.19 \ ln^{1/6} A [ x^3 - 1.5(x_0+x_L)x^2 + \nonumber  \\
3x_0x_Lx] - \Big[\alpha_A(r) - \frac{1.08(A^{1/3} -1) } {ln(A+1)} \sqrt{x} \Big] e^{-x^{2}/x_0^2},
\label{eq:hijing_shadow}
\end{align}

 where $ \alpha_A(r) = 0.1 ( A^{1/3} -1 )\frac{4}{3} \sqrt{1-r^2/R_A^2}. $ \\ 
Here r is the transverse distance of the interacting nucleon from the centre of the nucleus,  $R_A$ is the 
radius of the nucleus and value of  $x_0=0.1$ and $x_L =0.7$.  The significant nuclear dependence term is 
proportional to $\alpha_A(r) $, which determines the shadowing for $x<x_0$. Other terms gives small {\it A}-dependent nuclear effect on the structure function for $x>x_L$.



\subsection{The AMPT model}
\label{Ampt}

A Multi Phase Transport model (AMPT) \cite{ampt} use spatial and momentum distribution of minijet 
partons and  strings from HIJING . In the string melting approach (used in this study), without any further interaction these get converted into partons 
according to the flavor and spin structures of their valence quarks. Zhang's Parton Cascade (ZPC) 
model \cite{zpc} deal with interaction of such produced partons. The model follows 
Boltzmann equation, where the differential cross-section for leading order 
two body partonic interaction/scattering for eg. $gg\rightarrow gg$ is given as follows:
\begin{equation}
\frac{d\sigma_{gg}}{dt} \approx  \frac{9\pi \alpha_s^2}{2(t - \mu^2)^2},
\end{equation}
where $\alpha_s$ is the strong coupling constant, t is the standard Mandelstam variables for 
momentum transfer and $\mu$ is the screening mass of partonic matter.


Impact parameter dependent shadowing effect is present in AMPT, which is similar to that of HIJING (Eq \ref{eq:hijing_shadow}). In addition to this, AMPT includes $Q^2$ and flavor independent parameterization.

 AMPT uses a quark coalescence method to form hadrons once partons stop interacting, where two nearby 
quarks form a meson and three form a baryon. Hadronic interaction is modelled by ART model. 
This includes baryon-baryon, baryon-meson, and meson-meson elastic and inelastic scatterings.


\subsection{The NLO model}
\label{nlo}

The next-to-leading order, NLO-pQCD(MNR)~\cite{MNR,MNR_a} model is used in this 
work to produce \textit{c$\bar{\textit c}$} pair 
cross-sections in $p$+$p$ collisions at the next-to-leading order level. 
In the present work, we have used the calculations to produce $D\overline{D}$ and $Dh$ 
azimuthal angular difference, $\Delta\phi$=$\phi_D -\phi_{\overline{D}/h}$ correlation 
for $p$+$p$ collisions at $\sqrt{s}$ = 7 TeV and $p$+Pb at $\sqrt{s_{NN}}$ = 5.02 TeV.
The $p+$Pb scattering also includes shadowing~\cite{shadow2,shadow2_a} 
effects as one of the initial cold nuclear effects . Let us now move to a 
brief description of the calculations:

The correlation, '$C$' of heavy quarks produced in $pp$ collisions is defined in general as\cite{jamil,corr2}:
\begin{equation}
E_1 E_2\frac{d\sigma}{d^{3}p_1 d^{3}p_2}=
\frac{d\sigma}{dy_1 dy_2 d^{2}p_{T_1} d^{2}p_{T_2}}
=C~,
\label{ini1}
\end{equation}
where $y_1$ and $y_2$ are the rapidities of final quark-anti-quark system and 
$\bf{p_T}_i$ are their transverse momenta.

At the leading order, the differential cross-section for the charm correlation from proton-proton 
collision can be written as:
\begin{equation}
C_{LO}=\frac{d\sigma}{d^{2}p_{T}dy_1 dy_2}
\delta{(\bf{p_T}_1+\bf{p_T}_2)}~.
\end{equation}

In the above $C_{LO}$, it is assumed that

 $\bf{p_T}_1=\bf{-p_T}_2=\bf{p_T}$.

In the above, Eqn.~\ref{ini1}
\begin{eqnarray}
 \frac{d\sigma}{dy_1 dy_2 d^{2}p_{T_1} d^{2}p_{T_2}}  =   2 x_{a}x_{b}\sum_{ij}  \bigg[ f^{(a)}_{i}(x_{a},Q^{2})f_{j}^{(b)}(x_{b},Q^{2}) \frac {d\hat{\sigma}_{ij}(\hat{s},\hat{t},\hat{u})} {d\hat{t}}   \nonumber  \\
+ f_{j}^{(a)}(x_{a},Q^{2})f_{i}^{(b)}(x_{b},Q^{2}) \frac{d\hat{\sigma}_{ij}(\hat{s},\hat{u},\hat{t})}{d\hat{t}} \bigg] /(1+\delta_{ij}) \ ,
\label{ini2}
\end{eqnarray}

where $x_{a} $ and $x_{b} $ are the fractions of the momenta carried by the partons 
from their interacting parent hadrons, and are defined as,
\begin{equation}
x_{a}=\frac{M_{T}}{\sqrt{s}}(e^{y_1}+e^{y_2});~ x_{b}=\frac{M_{T}}{\sqrt{s}}(e^{-y_1}+e^{-y_2})~,
\end{equation}
where $M_{T} $ is the transverse mass, $\sqrt{m_{Q}^{2}+p_{T}^{2}}$, of the produced heavy quark.
The subscripts $i$ and $j$ denote the interacting partons, and $f_{i}$ and $f_{j}$ 
are the partonic distribution functions for the nucleons. 

We have used CTEQ6.6 \cite{cteq66} structure function as obtained using LHAPDF library for $p$+$p$ system and 
added EPS09 \cite{eps09} shadowing parameterization, 
to incorporate the initial nuclear effects on the parton densities for $p$+Pb system.

 The differential cross-section for partonic interactions, $d\hat{\sigma}_{ij}/d\hat{t}$ 
is given by
\begin{equation}
\frac{d\hat{\sigma}_{ij}(\hat{s},\hat{t},\hat{u})}{d\hat{t}} = \frac{\left|M\right|^{2}}{16\pi\hat{s}^{2}} ,
\label{dsdt}
\end{equation}
where $\left|M\right|^{2}$ is the invariant amplitude for various 
partonic sub-processes both for leading order (LO) and next-to-leading order (NLO) processes. \\
The physical sub-processes included for the leading order, 
$\cal{O}$ $(\alpha_{s}^{2}) $ production of heavy quarks are
\begin{eqnarray}
 g+g  \rightarrow Q+\overline{Q} \, and \nonumber \\
 q+\bar{q}  \rightarrow Q+\overline{Q} \ . 
\end{eqnarray}
At next-to-leading order, $\cal{O}$ $(\alpha_{s}^{3})$ sub-processes 
included are as follows:
\begin{eqnarray}
g+g  \rightarrow Q+\overline{Q}+g \ , \nonumber\\
q+\bar{q} \rightarrow Q+\overline{Q}+g \ , \nonumber\\
g+q(\bar{q}) \rightarrow Q+\overline{Q}+q(\bar{q}) .
\end{eqnarray}

To discuss briefly, the radiation process (giving associated gluons) in the NLO scenario, we know that 
the basic kinematics,
\begin{eqnarray}
 k_1^{q/g} + k_2^{\bar{q}/g} =  p_1^Q + p_2^{\overline{Q}}\,\nonumber,
\end{eqnarray}
goes to
\begin{equation}
 k_1^{q/g} + k_2^{\bar{q}/g} =  p_1^Q + p_2^{\overline{Q}} + k^g,
\end{equation}
where $k_1$, $k_2$ are the four momenta of incoming partons, $p_1$, $p_2$ and $k$ are 
the four momenta of the final charm quark pair and its associated gluon, with 
$\hat{s}, \hat{t}, \hat{u}$ mandelstam variables, for calculating invariant amplitude $|M|^2$ of 
2$\rightarrow$2 process has two additional terms $\hat{t}_k, \hat{u}_k$, containing four momentum, $k$. 
This leads to the logarithmic dependencies in the leading-order diagrams following Altarelli-Parisi 
formalism~\cite{MNR,AP}. The divergences in the cross-sections are controlled by the renormalization and factorisation 
parameters, $\mu_R$ and $\mu_F$ respectively.

Next we discuss re-scattering processes within the nucleus of $p+$Pb system.
A parton undergoes multiple hard scattering or a 
 nucleon instead undergoes multiple soft re-scattering within the cold nucleus in cases of $p$+A or A+A collisions. 
This is commonly referred as Cronin effects \cite{cronin1,accardi1,accardi1_a}. 
These re-scatterings may lead to momentum($k_T$) broadening of the 
interacting partons and change the final heavy quark spectrum. 
The details of our implementations of the calculations are taken from 
Ref. \cite{accardi2,sharma1,levai1, gyulassy1,gyulassy1_a}. 

We can now discuss briefly about the $k_T$ broadening. In the parton density function, 
\begin{equation}
f^{(a)}_{i}(x_{a},Q^{2},k_T^2)=f_{i}^{(a)}(x_{a},Q^{2}).g_{p/A}(k_T^2) \ , \\
\end{equation}
where \ $g_{p/A}(k_T^2)\propto exp[-k_T^2/\pi\,.\langle k_T^2\rangle_{pp/pA}]$  \nonumber 
and \ $\langle k_T^2\rangle_{pA}= \langle k_T^2\rangle_{pp} + \langle k_T^2\rangle_A$ . \nonumber 

The effective transverse momentum kick, 
$\langle k_T^2\rangle_{pA}$, following refs. \cite{accardi2, rsy1}
 is obtained by adding $\langle k_T^2\rangle_A$ to the intrinsic $\langle k_T^2\rangle_{pp}$. 
Our preliminary assumption of 
taking this summation however does not extrapolate $p+A$ system exactly to $p+p$ scenario. 
The $\langle k_T^2\rangle_A$ can be assumed as 
\begin{equation}
 \langle k_T^2\rangle_A = \delta^{2} . n.\ln \bigg( 1+\frac{p_T^2}{\delta^2/c} \bigg),
\end{equation}
where the parameters $\delta^2/c$, average squared momentum kick per scattering 
and $n=L_A/\lambda\,,L_A=4R_A/3$, average number of re-scattering, are taken from \cite{sharma1}.

With the implementation of the above features, we can next fragment the charm
momentum both from $p$+A and $p$+$p$
collisions into D-mesons, as
D-mesons data are readily verifiable from experiments. 
The fragmentation of the heavy quark $Q$ into the heavy-meson $H_M$ is
described by the Peterson fragmentation function $D_D(z)$ \cite{peterson1}. Similarly 
the associated gluon, $g$, is fragmented using global parametrization~\cite{bkk} following the
Binneweis and Kramer fragmentation function for gluons into $\pi, K, p$.



\section{Results and Discussions}
\label{result}
In this section we have shown our results on azimuthal correlation. $C(\Delta\phi)$ of $D$ mesons $(D^{0}, D^{+} \ and \ D^{*+}$ and their anti-particles) with their associated  hadrons $(\pi, k \ and \ p $ and their anti-particles) and $D$ mesons with their anti-partners, i.e. $\overline{D}$ mesons. 
We have used HIJING, AMPT and NLO-pQCD calculations 
to obtain the correlation spectra in azimuthal angles in the transverse plane. 
The models have been described in the 
previous sections. D meson rapidity is chosen as $-0.96<y_{D} (or \ y_{\bar{D}})<0.04$ for $p+$Pb system 
and $-0.5 <y_{D} (or \ y_{\bar{D}})< 0.5$ for $p+p$ system, where associated hadrons pseudo-rapidity is taken as $|\eta_{h}| < 0.8 $.  Difference in pseudo-rapidity window is taken as  $|\Delta\eta |=\eta_{h}(or \ \eta_{\bar{D}}) -\eta_D<$ 1.0.

\begin{figure}
\centering
 \resizebox{.6\textwidth}{!}{\includegraphics{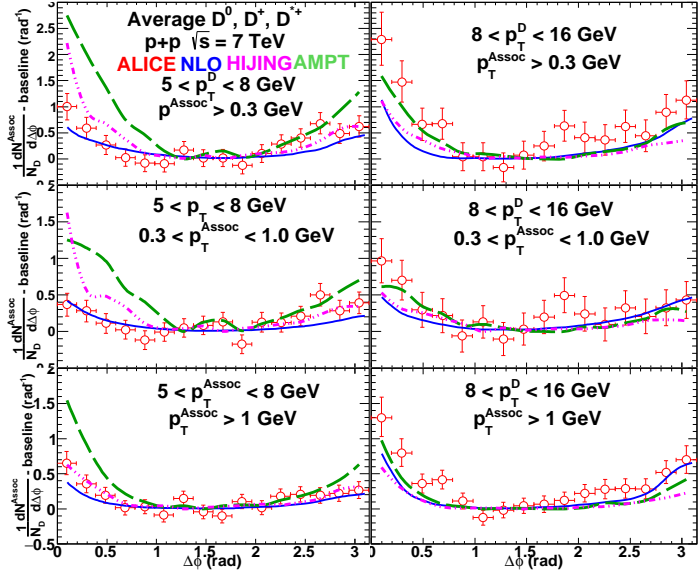}}
\caption{\small(Color online) Comparison of azimuthal correlation of D-mesons in $p+p$ $\sqrt{s}$ = 7 TeV. The left column represents $p_{T}$ of D-mesons with  $5 <  p_{T} < 8 $ GeV and the right column is   $8 <  p_{T} < 16$ GeV. The three rows represent, $p_{T}$ of charged hadrons (associated tracks) as $p_{T} > 0.3 $ GeV,  $0.3 <  p_{T} < 1$ GeV and  $p_{T} > 1$ GeV. Red circles represent ALICE data points \cite{alice_paper} . Green (dashed), magenta (dot-dashed) and blue(solid) lines represent AMPT, HIJING and NLO  results respectively.}
\label{fig:pp_dh}
\end{figure}

In the Fig. \ref{fig:pp_dh}, we have shown $C(\Delta\phi)$, correlation of $D$ meson with its associated hadron, 
for $p+p$ collisions at $\sqrt{s}$= 7 TeV. We have used some recent experimental data, 
two $p_T$ regions for D mesons, namely 
5.0 GeV $<p_{TD}<$ 8.0 GeV and 8.0 GeV$<p_{TD}<$ 16.0 GeV. Similarly, associated hadron momenta have been limited 
to $p_{Tassoc}>$ 0.3 GeV, 0.3 GeV $<p_{Tassoc}<$ 1.0 GeV, and $p_{Tassoc}>$ 1.0 GeV. 
The experimental data is taken from \cite{alice_paper}. The theoretical results show close matches with the experimental data. 
At the cut set, AMPT explains the data quite well within errors matching the trend and magnitude of the 
experimental points largely for 8.0 GeV$<p_{TD}<$ 16.0 GeV. However for 5.0 GeV $<p_{TD}<$ 8.0 GeV, AMPT overestimates the data for the near
peak. HIJING and NLO also are closer to the data but NLO appears to be more flat 
in the mid-azimuthal region. Also both HIJING and NLO do not have space-time evolution of charm quarks 
before fragmentation, there may be some extra gluons emitted during evolution which will add to the final $Dh$ 
spectrum. This may also contribute to the deviation of NLO results to an extent with the experimental data. 
Also the mismatch between NLO and AMPT as 
well as HIJING may be due to large radiations present in the calculations, which tend to push the correlated 
pair more towards near-side ($\Delta\phi = 0$). However in 8.0 GeV$<p_{TD}<$ 16.0 GeV, the model results are closer to each other as
we feel that substantial amounts of associated charged hadrons are not produced since the $D\bar{D}$ cross-
section itself becomes smaller at that region. 

\begin{figure}
\centering
 \resizebox{.6\textwidth}{!}{\includegraphics{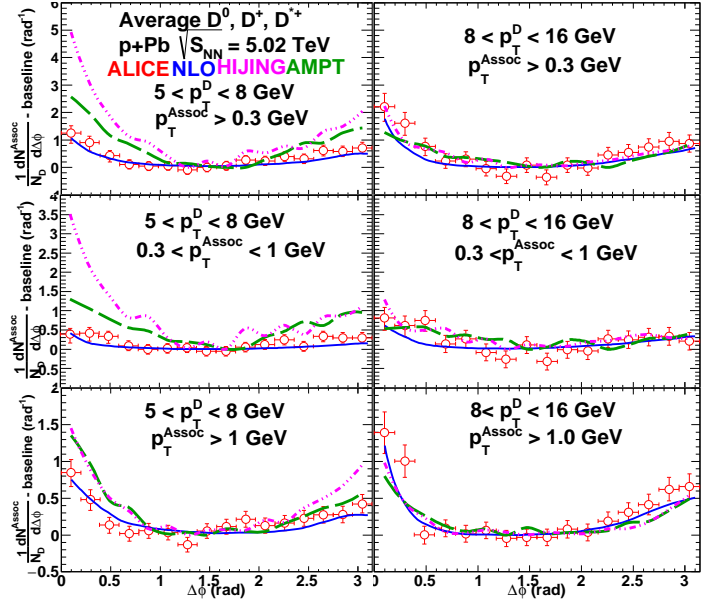}}
\caption{\small(Color online) Comparison of azimuthal correlation of D-mesons in $p$+Pb $\sqrt{s_{NN}}$ = 
5.02 TeV. The left column represents $p_{T}$ of D-mesons with  $5 <  p_{T} < 8 $ GeV and the right column is   $8 <  p_{T} < 16$ GeV. The three rows represent, $p_{T}$ of charged hadrons (associated tracks) as $p_{T} > 0.3 $ GeV,  $0.3 <  p_{T} < 1$ GeV and  $p_{T} > 1$ GeV. Red circles represent ALICE data points \cite{alice_paper} .  Green (dashed), magenta (dot-dashed) and blue(solid) lines represent AMPT, HIJING and NLO  results respectively.}
\label{fig:pPb_dh}
\end{figure}

In Fig. \ref{fig:pPb_dh},  we have shown $C(\Delta\phi)$, correlation of $D$ meson with its associated hadron, 
for $p$+Pb collisions at $\sqrt{s_{NN}}$= 5.02 TeV. 
The $p_T$ window used here is the same as that of last figure (Fig. \ref{fig:pp_dh}).
Both AMPT and NLO explain the experimental data well within errors. The curves show that 
near side correlation ($\Delta\phi$=0) is stronger with slightly higher peak than 
away side ($\Delta\phi$=$\pi$). Although in $Pb+Pb$, scenario the formation of hot and dense medium 
does alter the correlation due to medium modification of momenta of D mesons and their associated hadrons, 
the explanation of more near side peak in $p+Pb$ may be associated with multi parton scattering in the cold 
nucleus and due to shadowing effects, although such modification may 
be overshadowed by the effects of QGP in the heavy ion scenario. The theoretical calculations using 
NLO-pQCD, AMPT and HIJING are in better agreement with each other in this case.   
AMPT explains the result but overestimates  the data in the 5.0 GeV $<p_{TD}<$ 8.0 GeV region, mostly at 
the near-sided peak. We are currently looking into the reasons behind such deviations and 
will report in our future works.

\begin{figure}
\centering
 \resizebox{.6\textwidth}{!}{\includegraphics{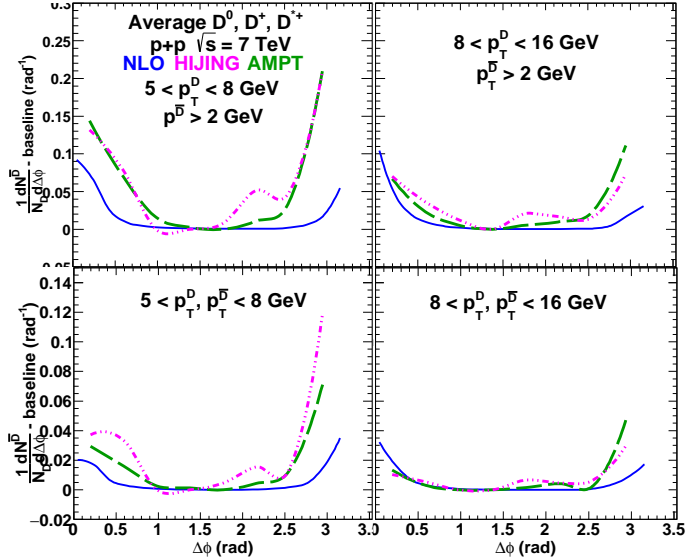}}
\caption{\small(Color online) Prediction of D-$\bar{D}$ azimuthal correlation in $p+p$ $\sqrt{s}$ = 7 TeV. The left column represents $p_{T}$ of D-mesons with  $5 <  p_{T} < 8$ GeV and the right column is   $8 <  p_{T} < 16$ GeV. The two rows represent, $p_{T}$ of $\bar{D}$ (associated tracks) as $p_{T} > 2$ GeV and in the same $p_{T}$ window as that of D-mesons. . Green (dashed), magenta (dot-dashed) and blue(solid) lines represent AMPT, HIJING and NLO  results respectively.}
\label{fig:pp_ddbar}
\end{figure}

In Fig. \ref{fig:pp_ddbar}, we plot $D\overline{D}$ azimuthal correlation for $p+p$ collisions at $\sqrt{s}$=7.0 TeV. 
The plots from NLO calculations show, for  $p_{T\bar{D}} > $2 GeV, both $p_{T}$ windows of D-mesons (5.0 GeV $<p_{TD}<$ 8.0 GeV and 8.0 GeV $<p_{TD}<$ 16.0 GeV) show a larger near side peak at $\Delta\phi$=0 and a smaller away side peak at $\Delta\phi$= $\pi$.  However when $p_{T}$ windows of D and $\overline{D}$ are same, then the peak is lightly away side for 5.0 GeV $<p_{TD,\bar{D}}<$ 8.0 GeV and for the case of 8.0 GeV $<p_{TD,\bar{D}}<$ 16.0 GeV the peak is slightly more near side. The middle region is flat or with little shape which shows that most of D meson pairs are either close to collinear or back-to-back correlated. 
 The HIJING and AMPT results show having larger away side peak for both $p_{T}$ windows and remains comparable. We would keep investigating and report on these differences in the future.

\begin{figure}
\centering
 \resizebox{.6\textwidth}{!}{\includegraphics{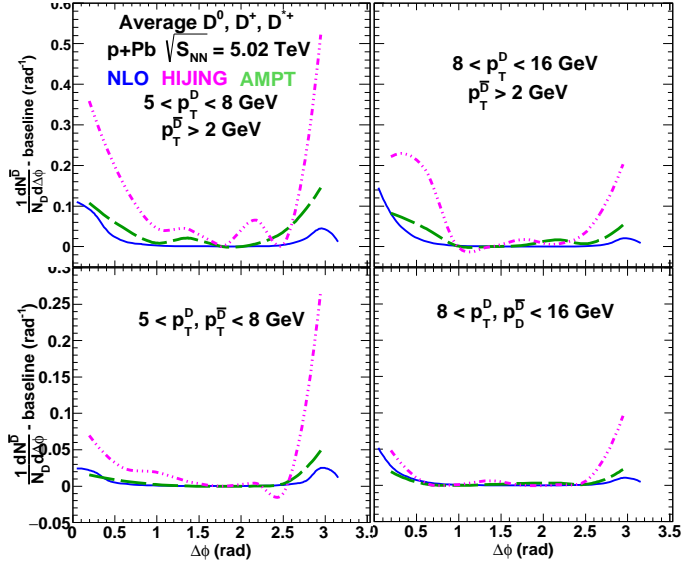}}
\caption{\small(Color online) Prediction of D-$\bar{D}$ azimuthal correlation in $p$+Pb $\sqrt{s_{NN}}$ = 5.02 TeV. The left column represents $p_{T}$ of D-mesons with  $5 <  p_{T} < 8$ GeV and the right column is   $8 <  p_{T} < 16$ GeV. The two rows represent $p_{T}$ of $\bar{D}$ (associated tracks) as $p_{T} > 2$ GeV and in the same $p_{T}$ window as that of D-mesons . Green (dashed), magenta (dot-dashed) and blue(solid) lines represent AMPT, HIJING and NLO  results respectively.}
\label{fig:pPb_ddbar}
\end{figure}

In Fig. \ref{fig:pPb_ddbar}, we plot $D\overline{D}$ azimuthal correlation for $p+Pb$ collisions at $\sqrt{s_{NN}}$=5.02 TeV. 
Here due to some effects of $k_T$ broadening (because of multi parton scattering) and also due to NLO effects, the distributions give steeper peaks at the near side for all the $p_T$ ranges except when $p_{T}$ ranges of D-$\bar{D}$ are both 5.0 GeV $<p_{TD}<$ 8.0 GeV. Similarly 
for HIJING and AMPT we have more away-sided peak like that of $p+p$ results, which shows more effect 
of radiation during space-time evolution of charm quarks but considerable CNM effects are also evident due to shadowing and multi partonic scattering from the p+Pb plot compared to p+p.

Overall, as in the case of QGP, hot and dense matter has overshadowing CNM effects to large extent, and
this $p+Pb$ study could give us an indirect view of effects of CNM phenomena 
choosing right $p_T$ windows. Since the effects of such 
CNM effects on azimuthal distribution in different $p_T$ regions are still under investigation, we will report 
some general implications in such studies in our future works.

\section{Conclusion}
\label{summary}
We have shown correlations of $D$ mesons with their associated hadrons, $h$ or with 
their anti-partner $\overline{D}$, in azimuthal angles in the transverse momentum plane. 
The models used here are AMPT, HIJING, and NLO-pQCD. The models agree with the experimental data but the disagreements are mostly at near side peak.\\  
NLO-pQCD does explain experimental data well in  some $p_{T}$ windows, but it underestimates data in other.  Absence of partonic evolution in NLO-pQCD might be one possible reason. The coalescence method of hadronisation in AMPT is producing bigger differences than fragmentation of HIJING particularly in the $5 <p _{T} < 8 $ GeV regions, while in $p_{T} > 8$ GeV, the results are similar. 
In the $5 <p _{T} < 8 $ region, a reversal of results from AMPT to HIJING between p+p to p+Pb system might be from excessive radiation in AMPT due to multi-parton interaction in cold nuclear matter and hence resulting in lowering of heavy quark spectra (we observed similar results in \cite{rsy1}).

Also some $p_T$ cuts do show effect of anisotropy in 
such two particle azimuthal angular difference distributions. 
These may be due to certain kinematical constraints or effects 
of CNM effects which are being currently investigated and would be 
discussed in detail in our future works. 

As for $D-\bar{D}$ plots, since no experimental data is yet available, the current results may 
serve as qualitative predictions while analysis by experimental groups will shed light on this topic 
and help us to constraint our models further.

Authors would like to thank  Somnath Kar of VECC, Kolkata, India for his fruitful comment and discussion.

%
%

\end{document}